\documentclass[twocolumn,showpacs,amsmath,amssymb]{revtex4}

\usepackage{graphicx}
\usepackage{dcolumn}
\usepackage{bm}
\usepackage{epsfig}
\usepackage{amsfonts}

\begin{document}


\title{Axiomatic approach to the cosmological constant}

\author{Christian Beck}
\affiliation{
School of Mathematical Sciences \\ Queen Mary, University of
London, Mile End Road, London E1 4NS, UK}
\email{c.beck@qmul.ac.uk}
\homepage{http://www.maths.qmul.ac.uk/~beck}

\date{\today}

\vspace{2cm}

\begin{abstract}
A theory of the cosmological constant $\Lambda$ is currently out
of reach. Still, one can start from a set of axioms that describe
the most desirable properties a cosmological constant should have.
This can be seen in certain analogy to the Khinchin axioms in
information theory, which fix the most desirable properties an
information measure should have and that ultimately lead to the
Shannon entropy as the fundamental information measure on which
statistical mechanics is based. Here we formulate a set of axioms
for the cosmological constant in close analogy to the Khinchin
axioms, formally replacing the dependency of the information
measure on probabilities of events by a dependency of the
cosmological constant on the fundamental constants of nature.
Evaluating this set of axioms one finally arrives at a formula
for the cosmological constant given by
$\Lambda=\frac{1}{\hbar^4}G^2 \left(
\frac{m_e}{\alpha_{el}}\right)^6$, where $G$ is the gravitational
constant, $m_e$ the electron mass, and $\alpha_{el}$ the low
energy limit of the fine structure constant. This formula is in
perfect agreement with current WMAP data. Our approach gives
physical meaning to the Eddington-Dirac large number hypothesis
and suggests that the observed value of the cosmological constant
is not at all unnatural.
\end{abstract}

\pacs{xxx}

\date{\today}          
\maketitle

\section{Introduction}

The cosmological constant problem is probably one of the most
fundamental problems in physics that so far has resisted any
attempt of solution \cite{weinberg}. When looking through the
large amount of literature on the cosmological constant $\Lambda$
and the associated cosmological constant problem, one statement is
found quite regularly: The observed value of the cosmological
constant (or dark energy density) that is suggested by WMAP and
other astronomical observations \cite{spergel, riess} is regarded
by most physicists to be rather unnatural and surprising, and some
people from the anthropic school even regard it to be
unexplainable. From a quantum field theory point of view one
would have a priori expected a value of vacuum energy density
$\rho_{vac}= \frac{c^4}{8\pi G} \Lambda$ given by something of
the order $m_{pl}^4$ (in units where $\hbar = c =1$), since the
Planck mass $m_{pl}$ is a suitable cutoff scale for vacuum
fluctuations where quantum field theory needs to be replaced by
something else. So for a quantum field theorist the observed
value of the cosmological constant is surprisingly small.
Astrophysicists, on the other hand, are facing a rather large
value of $\Lambda$ in the $\Lambda CDM$ model as compared to
observable mass densities. This means that the current
universe is dominated by vacuum energy, whereas a priori most
astrophysicists would have probably expected dark energy to play
a less pronounced role, so for them $\Lambda$ is surprisingly
large.
In supersymmetric theories, and in particular superstring theory,
the most natural value of the cosmological constant is zero, and
again it is not clear how to obtain a small positive value at the
current time by a 'natural' mechanism. This has led to models
based on anthropic considerations, which give up the idea of a
single universe and regard $\Lambda$ as a random variable whose
value (by construction of the anthropic argument) can never be
explained, just probabilistic statements can be given for an
ensemble of `multiverses' \cite{carr}.

Given all these controversies and mysteries surrounding the
cosmological constant, it is perhaps worth to go back to the
basics and to ask ourselves how natural or unnatural the observed
value really is. Since the development of a full theory of the
cosmological constant is currently out of reach, we will start by
formulating a set of axioms that describe the most desirable
properties a cosmological constant should have. This can be seen
in analogy to the set of Khinchin axioms \cite{khinchin, abe} in
information theory that describe the most desirable properties an
information measure should have. It is well-known that from the
Khinchin axioms one can uniquely derive the Shannon entropy, on
which the entire mechanism of statistical mechanics is founded
(see any textbook on the subject, e.g.\ \cite{beck-schloegl}).
Similarly, we will formulate suitable axioms for a cosmological
constant. The principal idea underlying this approach is that
ultimately the cosmological constant is expected to be part of a
unified theory of quantum gravity and the standard model of
eletroweak and strong interactions. It will thus potentially
depend on fundamental constants of nature. The axioms that we
formulate deal with possible dependencies on these fundamental
parameters.

Roughly speaking, the physical contents of these axioms is as
follows: The cosmological constant should only depend on
fundamental parameters of nature (rather than irrelevant
parameters), it should be bounded from below, it should depend on
the relevant fundamental constants in the simplest possible way,
and the dependence should be such that one obtains scale
invariance of the universe under suitable transformations of the
fundamental parameters that leave the physics invariant. We will
point out that these 4 axioms for the cosmological constant are
very similar in style to the 4 Khinchin axioms that ultimately
underlie the foundations of statistical mechanics. Amazingly, out
of the 4 axioms a formula for $\Lambda$ can be derived that is in
excellent agreement with current observations. This formula is
given by
\begin{equation}
\Lambda = \frac{1}{\hbar^4} G^2 \left( \frac{m_e}{\alpha_{el}}
\right)^6, \label{1111}
\end{equation}
where $G$ is the gravitational constant, $m_e$ the electron mass,
and $\alpha_{el}$ the low-energy limit of the fine structure
constant.

It turns out that the result (\ref{1111}) of our derivation can
be interpreted as a particular form of the Eddington-Dirac large
number hypothesis, connecting cosmological parameters and
fundamental constants \cite{eddington, dirac, not1, not2,
dematos}, in a form previously advocated by Nottale
\cite{not1,not2}, based on previous work by Zeldovich \cite{zel}.
Still our derivation is very different from Nottale's original
approach, since we do not need any assumption on 'scale
relativity' \cite{not2}. Rather, our method is much more related
to an information-theoretic approach (similar as in statistical
mechanics) and gives new physical meaning to this kind of
approach. The above value of the cosmological constant is singled
out as a kind of optimum value that is consistent with the axioms.
The validity of formula (\ref{1111}) has also been independently
conjectured in a recent paper by Boehmer and Harko \cite{boehmer}.

The vacuum energy density (dark energy density with
equation of state $w=-1$)
that follows from our axiomatic approach is given by
\begin{equation}
\rho_{vac}=\frac{c^4}{8\pi G} \Lambda =\frac{1}{8\pi}
\frac{c^4}{\hbar^4}G \left( \frac{m_e}{\alpha_{el}} \right)^6.
\label{rho-vac}
\end{equation}
According to the 4 axioms that we will formulate in the following
sections, eq.~(\ref{rho-vac}) yields a kind of an optimum value of
the vacuum energy in the universe, according to criteria set out
by the axioms. Numerically, this formula yields the prediction
\begin{equation}
\rho_{vac}=(4.0961 \pm 0.0006) GeV/m^3 \label{rhovalue}
\end{equation}
The current astronomical measurements provide evidence for a dark
energy density of about
\begin{equation}
\rho_{dark}=(3.9 \pm 0.4) GeV/m^3.
\end{equation}
We thus conclude that the observed value of the cosmological
constant is not at all unnatural, but derivable from a set of
suitable axioms that make physical sense.

This paper is organized as follows. In section 2 we briefly
recall the Khinchin axioms, in oder to make this paper
self-contained for readers that do not have an information theory
background. In section 3 we formulate our 4 axioms for the
cosmological constant and point out the analogy with the Khinchin
axioms. In section 4 we derive the above formula for $\rho_{vac}$
from the axioms. In section 5 we point out that our axiomatic
approach gives physical meaning to the Eddington-Dirac large
number conjecture. Our concluding remarks are given in section 6.

Throughout this paper our notion of $\rho_{vac}$
means the observable (physically relevant) vacuum
energy density, which should be distinguished from any bare
(unmeasurable) contributions, see e.g. \cite{beck-mackey1}
for a discussion of this subtlety.

\section{The Khinchin axioms}

Khinchin \cite{khinchin} formulated four axioms that describe the
most desirable properties an information measure $I$ should have.
These 4 axioms uniquely fix the functional form of the Shannon
information and are extremely important for the mathematical
foundations of statistical mechanics. Let us recall these 4
axioms as well as their physical meaning. Later, we will
formulate analogous axioms for the cosmological constant.

{\bf A1 'fundamentality'}
\begin{equation}
I=I(p_1,\cdots ,p_W)
\end{equation}
That is to say, an information measure $I$ only depends on the
probabilities $p_i$ of the events under consideration
($i=1,\ldots, W$) and nothing else. It should not depend on any
other irrelevant or non-fundamental quantities.

{\bf A2 'boundedness'}
\begin{equation}
I(W^{-1},\ldots ,W^{-1})\leq  I(p_1,\cdots ,p_W)
\end{equation}
This means the information measure $I$ takes on an absolute
minimum for the uniform distribution $(W^{-1},\ldots ,W^{-1})$,
every other probability distribution has an information contents
that is larger or equal to that of the uniform distribution.
Clearly, this implies there is a lower bound for $I$. We may thus
call this the 'axiom of boundedness'.

{\bf A3 'simplicity'}
\begin{equation}
I(p_1,\ldots,p_W)=I(p_1,\ldots,p_W,0)
\end{equation}
This means the information measure $I$ should not change if the
sample set of events is enlarged by another event that has
probability zero. Clearly, one can make the description of a
given model as complicated as one like, but this axiom advocates
the simplest description, where irrelevant events with probability
zero are excluded. We may call this the 'axiom of simplicity'.

{\bf A4 'invariance'}
\begin{equation}
I(\{p_{ij}^{I, II}\})=I(\{p_i^I\})+\sum_i p_i^I  I(\{p^{II}(j|i)
\})
\end{equation}
This axiom is slightly more complicated and requires a longer
explanation. The axiom deals with the composition of two systems I
and II (not necessarily independent). The probabilities of the
first system are $p_i^I$, those of the second system are
$p_j^{II}$. The joint system $I,II$ is described by the joint
probabilities $p_{ij}^{I, II}=p_i^Ip^{II}(j|i)$, where
$p^{II}(j|i)$ is the conditional probability of event $j$ in
system II under the condition that event $i$ has occurred in
system $I$. $I(\{p^{II}(j|i)\})$ is the conditional information of
system II formed with the conditional probabilities $p^{II}(j|i)$,
i.e.\ under the condition that system I is in state $i$.

The meaning of the above axiom is that it postulates that the
information measure should be independent of the way the
information is collected. We can first collect the information in
the subsystem II, assuming a given event $i$ in system I, and then
sum the result over all possible events $i$ in system I, weighting
with the probabilities $p_i^I$.

For the special case that system I and II are independent the
probability of the joint system factorizes as
\begin{equation}
p_{ij}^{I,II}=p_i^Ip_j^{II},
\end{equation}
and only in this case, axiom 4 reduces to the rule of additivity
of information for independent subsystems:
\begin{equation}
I(\{ p_{ij}^{I,II}\})=I(\{p_i^I\})+I(\{p_j^{II}\}) \label{indep}
\end{equation}

Apparently, the above axiom deals with the fact that there is some
invariance of the information measure if a description in terms of
other probabilities (namely the conditional ones) associated with
other subsystems is chosen. In a more abstract way we may write
\begin{equation}
I (\{ \tilde{p_i} \})= \tilde{I} (\{ p_i \})
\end{equation}
meaning that there is a suitable scale transformation (denoted by
$\tilde{\;}$) in the space of probabilities and information
measures that leaves the physical contents invariant. We may thus
call the 4th axiom the 'axiom of invariance'.

One can now easily show that the functional form of the Shannon
information
\begin{equation}
I(p_1, \ldots p_W) =\sum_i p_i \log p_i
\end{equation}
follows uniquely from axioms A1--A4 (see any textbook on this
topic, for example \cite{beck-schloegl}). The fourth axiom
actually is the most important one. While there are many different
information measures satisfying A1--A3, the specific form of the
Shannon information is crucially determined by A4 \cite{abe}.

\section{Axioms for the cosmological constant}

We now proceed to axioms for the cosmological constant, which
will turn out to share a certain analogy with the Khinchin
axioms. The role of the information measure is now formally
played by $\Lambda$, and the dependence of $I$ on the
probabilities $p_i$ is replaced by the dependence of $\Lambda$ on
the fundamental constants of nature, such as coupling constants
$\alpha_i$, masses $m_i$, and mixing angles $s_i$. Let us now
formulate the following set of axioms B1--B4:

{\bf B1 'fundamentality'}

$\Lambda =\Lambda (\{ \alpha_i \}, \{ m_i \}, \{s_i \})$

The cosmological constant depends on fundamental constants of
nature only. There is no dependence on non-fundamental parameters.

{\bf B2 'boundedness'}

$0 <\Lambda$

The cosmological constant is bounded from below. The trivial
solution $\Lambda =0$ is not allowed.

{\bf B3 'simplicity'}

$\Lambda (\{ \alpha_i \}, \{ m_i \}, \{s_i \}) =\Lambda (\{
\alpha_i \}, \{ m_i \}, \{s_i \}, \{ c_i \})$

The cosmological constant is given by the simplest possible
formula consistent with the other axioms, avoiding irrelevant
non-universal prefactors or dependencies on irrelevant parameters
$c_i$.

{\bf B4 'invariance'}

$\Lambda (\{ \tilde{\alpha_i} \}, \{ \tilde{m_i} \}, \{\tilde{s_i}
\} )= \tilde{\Lambda} (\{ \alpha_i \}, \{ m_i \}, \{s_i \}) $

A cosmological constant formed with potentially different values
of fundamental parameters leaves the
large-scale physics of the universe scale invariant.

As we can see, axiom B1 is basically the same as Khinchin axiom
A1, formally replacing the information measure $I$ by the
cosmological constant $\Lambda$ and the dependence on the
probabilities $p_i$ of events by the dependence of $\Lambda$ on
the fundamental constants of nature. Axiom B2, just like axiom A2,
states that there is a lower bound on the cosmological constant,
which in our case is taken to guarantee that $\Lambda$ is bounded
away from zero, thus excluding the trivial solution. In contrast
to A2, we need a strict inequality in B2. Axiom B3, just like A3,
advocates that the simplest possible description is the
physically relevant one. Whereas A3 excludes irrelevant events
(those with probability 0), in a similar way B3 excludes
irrelevant non-fundamental constants from influencing $\Lambda$.
Finally, the most important and restrictive axiom is axiom B4.
Similar as A4, which yields the total information $I$ obtained by
using different types of probabilities in different subsystems, B4
deals with a description of the cosmological constant using
potentially different constants of nature in a different
universe. Similar to A4, B4 postulates that the large-scale
physics should be scale invariant under this transformation
process of fundamental parameters. In other words, in relative
terms the effect of the cosmological constant as compared to
other interaction strengths should be unchanged.

\section{Derivation of the cosmological constant from the axioms}

From B2 it follows that the cosmological constant has a nontrivial
positive value. From B1 and B3 it follows that this value can be
written as the simplest possible combination of relevant
fundamental constants of nature that is consistent with B4.

As for the Khinchin axioms, the 4th axiom for the cosmological
constant is also the most restrictive one. This axiom is dealing
with the physical effects that the cosmological constant has in
an evolving universe and puts it into relation with the other
interactions. B4 says that the large scale physical effect of the
cosmological constant should be invariant if relevant fundamental
parameters change to different values. In other words, the {\em
relative} effect of the cosmological constant as compared to the
other relevant large-scale physics should be the same under
fundamental parameters transformations. We note that the
cosmological constant acts on a large scale and hence it is
unlikely that it is influenced by the fundamental parameters of
strong and weak interactions, which are relevant for short-range
physics only. If at all, it should be influenced by interactions
that have a long range: gravity and electromagnetism.

Of physical meaning for the evolution of the universe is really
the vacuum energy density $\rho_{vac}=\frac{c^4}{8\pi G} \Lambda$
associated with $\Lambda$. In the following we will derive in
three steps the simplest form of vacuum energy density that is
consistent with axiom B4. In the first step, we apply B4 to
gravitational physics. In the second step, a dimensional argument
is given. Finally, in the third step we take into account
electromagnetic interaction processes. Putting together all 3
steps we finally end up with a concrete formula for $\rho_{vac}$.

\subsection{Step 1: Gravitational scale invariance}

Consider an arbitrary spatial volume $V$ of the universe at an
arbitrary time, which (at a late stage) may contain galaxies,
stars, dust etc. We denote the point masses contained in this
volume by $m_i$, $i=1,2,3,\ldots$ The total gravitational energy
density in this volume $V$ is
\begin{equation}
\rho_G=-\frac{G}{V} \sum_{i,j} \frac{m_im_j}{r_{ij}},
\label{rho-G}
\end{equation}
where $r_{ij}$ is the distance between masses $m_i$ and $m_j$ and
the sum runs over all indices $i,j$ with $i>j$. At the same time
there is also constant vacuum energy density $\rho_{vac}$ in this
region of the universe. Now consider a fundamental parameter
transformation in the universe which gives the gravitational
constant $G$ a new value. We keep all distances $r_{ij}$ and masses
$m_i$ as they are but formally change the gravitational constant
\begin{equation}
G \to \Gamma G. \label{Gamma}
\end{equation}
For example, $\Gamma =2$. Then the gravitational energy density
$\rho_G$ also doubles:
\begin{equation}
\rho_G \to \Gamma \rho_G
\end{equation}
Keeping the original vacuum energy density $\rho_{vac}$ would
clearly change the relative size of $\rho_G$ and
$\rho_{vac}$. Scale invariance of the universe as a whole is only
achieved if at the same time $\rho_{vac}$ is also transformed as
\begin{equation}
\rho_{vac} \to \Gamma \rho_{vac}
\end{equation}
Only in this case the ratio of gravitational energy density and
vacuum energy density stays the same and the universe remains
scale invariant. In other words, the postulate B4 of scale
invariance of the universe under the fundamental parameter
transformation (\ref{Gamma}) implies that the vacuum energy
density must be proportional to $G$:
\begin{equation}
\rho_{vac} \sim G X\label{1}
\end{equation}
Here $X$ denotes something that is so far unknown but independent
of the gravitational constant $G$. Eq.~(\ref{1}) already
represents an important step which strongly restricts the possible
dependencies of $\rho_{vac}$ on the fundamental constants of
nature, as we shall see in the following.

\subsection{Step 2: A dimensional argument}

Let us now discuss in more detail the form of the quantity $X$ in
eq.~(\ref{1}). For dimensionality reasons, in units where $\hbar
=c =1$, we must multiply $G=1/m_{pl}^2$ by some mass scale to the
power 6 in order to get an energy density, which in these units
has the dimension mass to the power 4. If we allow for an
arbitrary dimensionless proportionality factor $A$, we may choose
this reference mass scale to be the electron mass $m_e$, which
surely is a fundamental constant, and write
\begin{equation}
\rho_{vac}= A G m_e^6
\end{equation}
We now return to units where $\hbar$ and $c$ are measured in SI
units. In this case a prefactor $c^4/\hbar^4$ is needed to give
$\rho_{vac}$ the correct dimension of an energy per volume:
\begin{equation}
\rho_{vac}= A \frac{c^4}{\hbar^4} G m_e^6
\end{equation}
The arbitrary dimensionless proportionality factor $A$ can be
expressed rather arbitrarily in terms of the fine structure
constant $\alpha_{el}$ as $A=\frac{1}{8\pi} \alpha_{el}^\eta$ if
we allow $\eta$ to be an arbitrary real number. Hence
\begin{equation}
\rho_{vac}= \frac{1}{8 \pi} \frac{c^4}{\hbar^4} G m_e^6
\alpha_{el}^\eta \label{eta}
\end{equation}
Since $\eta$ is arbitrarily chosen, not very much is gained so
far. However, we will now show in the third and final step that
axioms B1,B3, and B4 applied to electromagnetic interaction
processes in the early universe will fix the so far arbitrary
real number $\eta$.

\subsection{Step 3: Electromagnetic scale invariance}

After recombination, almost all free charges in the universe
vanish due to the formation of atoms which are electrically
neutral on a large scale. Hence, in contrast to gravitational
interaction, electromagnetic interaction is irrelevant on large
scales after recombination. The large-scale physics of the
universe, however, is dominated by electromagnetic interaction
processes {\em before} recombination, where lots of free charges
exist that interact with photons. The relevant electromagnetic
scattering process before recombination is Thomson scattering.
The total cross section $\sigma_T$ for Thomson scattering of
particles of mass $m$ and charge $Q$ is given by
\begin{equation}
\sigma_T= \frac{8\pi}{3} \left( \alpha_{el} Q^2
\frac{\hbar}{mc}\right)^2.
\end{equation}
Note that this cross section is dominated by the lightest charged
particles, i.e. electrons (charge $Q=-1$, mass $m=m_e$). Compared
to electrons, Thomson scattering of heavy particles such as $\mu,
\tau$ or protons is clearly negligible in the early universe, due
to the higher mass involved. For electrons, we may write
\begin{equation}
\sigma_T = \frac{8\pi}{3} r_e^2
\end{equation}
where
\begin{equation}
r_e= \alpha_{el} \frac{\hbar}{m_e c}
\end{equation}
is the classical electron radius. Thomson scattering of electrons
is the most important scattering process in the early universe,
and is thought to be responsible for the linear polarization of
the cosmic microwave background. It dominates the large-scale
physics of the early universe and hence is {\em the}
electromagnetic process that, if any, should be describable in a
scale invariant way relative to the energy scale set by the
cosmological constant, according to axiom B4.

Let us now consider a fundamental parameter transformation of the
following form:

\begin{eqnarray}
\alpha_{el} &\to& \Gamma \alpha_{el} \\
m_e & \to &\Gamma m_e
\end{eqnarray}
For example, $\Gamma =2$. The above simulataneous transformation
of the electromagnetic parameters $\alpha_{el}$ and $m_e$ leaves
the physics of Thomson scattering invariant, since $\sigma_T$ only
depends on the ratio $\alpha_{el}/m_e$. Relative to this, we want
the cosmological constant to stay invariant as well under the
above transformation, because otherwise the relative size of the
inverse cosmological constant as compared to the Thomson cross
section would change, thus violating axiom B4, which requires
scale invariance of the universe as a whole (note that
$\Lambda^{-1}$ and $\sigma_T$ have the same dimension, namely
length squared, hence the two quantities are directly comparable).
Invariance under the above
simultaneous transformation with $\Gamma$ means that $\Lambda$,
respectively $\rho_{vac}$, must be a function of the ratio
$m_e/\alpha_{el}$ rather than a function of $m_e$ and
$\alpha_{el}$ on their own. But this fixes the parameter $\eta$
in eq.~(\ref{eta}) to be $\eta =-6$. We thus arrive at the final
result
\begin{equation}
\rho_{vac}= \frac{1}{8 \pi} \frac{c^4}{\hbar^4} G \left(
\frac{m_e}{\alpha_{el}} \right)^6. \label{eta6}
\end{equation}



One remark is still at order. The factor $1/8\pi$ in
eq.~(\ref{eta6}) was introduced by convention and it could still
be in principle something else. However, here axiom B1 and B3
help which state that the cosmological constant should be a
function of the fundamental constants of nature only, in the
simplest possible way, and avoiding arbitrary prefactors for
which there is no physical reason. The prefactor $1/8 \pi$ for
the vacuum energy $\rho_{vac}$ in eq.~(\ref{eta6}) is precisely
chosen in such a way that the cosmological constant
\begin{equation}
\Lambda = \frac{8 \pi G}{c^4} \rho_{vac} = \frac{1}{\hbar^4} G^2
\left( \frac{m_e}{\alpha_{el}} \right)^6 \label{Lam}
\end{equation}
does not have any prefactor. So this choice is indeed the
simplest possible one, in agreement with axiom B3 of simplicity.

\section{Connection with Eddington-Dirac large number hypothesis}

Using $G= \hbar c/ m_{pl}^2$ one can easily check that
eq.~(\ref{Lam}) can be written in the equivalent dimensionless form
\begin{equation}
\alpha_{el} \frac{m_{pl}}{m_e} = \left(
\frac{\Lambda^{-1/2}}{l_{pl}} \right)^{\frac{1}{3}},
\label{edd-dirac}
\end{equation}
where $l_{pl}=\hbar/m_{pl}c$ is the Planck length. This can be
regarded as an Eddington-Dirac large number hypothesis, since the
equation connects cosmological parameters with standard model
parameters \cite{eddington, dirac, siv}. Eq.~(\ref{edd-dirac}) has been
written down previously by Nottale \cite{not1}.
On both sides of the above equation one has two very large
numbers. So far these types of large-number relations have often
been regarded as being just some type of numerical coincidence.
But here we see that the occurence of such large numbers that
coincide is not at all unnatural. Rather, all this follows in a
rather straightforward way from our set of axioms.

One may ask why to take the power 1/3 in the above equation and
not some other power. If no theory is available, then this power
could in principle be chosen in an arbitrary way. However, the
power 1/3 follows in a stringent and logically consistent way out
of our axiomatic approach. In fact, the power 1/3 in
eq.~(\ref{edd-dirac}) is equivalent to the power 6 of $m_e$ in
eq.~(\ref{eta6}), and this power was derived in the previous
section from the postulate of gravitational scale invariance and
using our dimensional argument. We see that the above
Eddington-Dirac large number relation arises in quite a natural
way out of a set of axioms B1---B4 that do make physical sense
and that in a way describe the most desirable properties a
cosmological constant should have when compared with the other
relevant processes in an evolving universe. In this way our
axiomatic approach presented here has given physical meaning to
this relation, and makes it plausible that there is more to this
relation than just a numerical coincidence of some large numbers.

An interesting aspect is the fact that the
relation (\ref{Lam}) uses fundamental parameters such as $\hbar,
G, m_e, \alpha_{el}$ that are all known with very high precision.
Thus this relation  allows for an extremely precise prediction of
the value of the cosmological constant $\Lambda$, by far more
precise than the present cosmological observations can confirm:
\begin{equation}
\Lambda =(1.36284\pm 0.00028)\cdot 10^{-52} m^{-2}
\end{equation}
The above value can be used to precisely fix the
relevant energy scale in other, more advanced microscopic models
of dark energy (e.g. \cite{beck-prd, beck-mackey2}).

\section{Conclusion}

In this paper we have formulated a set of axioms that in a sense
describe the most desirable physical properties a cosmological
constant should have. This set of axioms can be seen in analogy
to the set of Khinchin axioms, which describe the most desirable
properties an information measure should have. The Khinchin axioms
uniquely lead to the Shannon information measure, which lies at
the root of the mathematical foundations of statistical mechanics
and thermodynamics. In a similar way, our set of axioms leads to
a formula for the cosmological constant that is equivalent to a
particular type of Eddington-Dirac large number hypothesis, in a
form previously advocated by Nottale. Our approach gives physical
meaning to this formula, which so far was only regarded to be a
numerical coincidence. The agreement of our derived formula with
the WMAP measurements of dark energy density in the universe is
amazing, given the fact that {\em a priori} our set of physically
reasonable axioms does not know anything about supernovae
measurements and could have led to a completely different vacuum
energy density prediction, different from the observed one by many
orders of magnitude.

The fact that techniques from information theory (Khinchin-like
axioms for the cosmological constant) lead to apparently successful
predictions of the numerical value of $\Lambda$  opens up new ways to deal with the cosmological
constant problem. In fact, tools from information theory and
thermodynamics have proved to be useful for gravitational physics
in the past as well, the main example being the thermodynamics of
black holes. Our approach here suggests that similar information
theoretic tools can be useful to derive concrete predictions for the
observed dark energy density in the universe, i.e. eq.~(\ref{rhovalue}).

Our approach certainly does not provide any microscopic theory
underlying the cosmological constant. This must be the task of
future work---our work here is just based on information-like
techniques as an effective theory. However, again we are reminded
of the foundations of statistical mechanics and thermodynamics:
These theories work perfectly, based on maximizing the Shannon
entropy subject to suitable constraints. The Shannon entropy is
identified with the physical Boltzmann-Gibbs entropy and its
functional form directly follows from the Khinchin axioms A1--A4.
Similarly, the cosmological constant $\Lambda (\{ \alpha_i \}, \{
m_i \}, \{ s_i \})$ in this paper is identified as the source of
measurable dark energy in the universe and its functional
dependence on the fundamental parameters follows from the axioms
B1--B4.

So far nobody has succeeded to provide a rigorous derivation of
statistical mechanics starting from the underlying microscopic
equations. Rather, one usually starts from the Khinchin axioms,
gets from them  the Shannon entropy and then proceeds to
statistical mechanics and thermodynamics by optimization of the
Shannon entropy. This method works perfectly but a deeper
understanding of the microscopic foundations of statistical
mechanics is not provided. Maybe the cosmological constant awaits
a similar fate: Its numerical value (the observed dark energy
density in the universe) can be made plausible using the
Khinchin-like axioms B1--B4 introduced in this paper, but the
microscopic quantum field theoretical foundations of the
cosmological constant are not elucidated by this approach. The
ultimate microscopic theory of a cosmological constant as embedded
in a theory of quantum gravity still has to be developed.

\end{document}